# Can We Trust AI Agents in the Supermarket? Sugar Content Inference from Product Images

Jose Berengueres, IEEE member
School of Computing & AI,
Nazarbayev University, Astana,
Kazakhstan
jose.berengueres@nu.edu.kz

***Abstract*—Nutritional labels are legally permitted to appear in very small print, reducing real-world readability and encouraging consumers to rely on "AI nutrition lens" and vision-capable conversational agents for dietary guidance. We evaluate whether such AI-mediated advice can meaningfully substitute for regulated labeling using a bounded, verifiable task: inferring which of two packaged foods contains less sugar from front-of-pack images alone. A Two-Alternative Forced Choice game was used to evaluate AI agent systems across four national supermarket contexts: Sweden, the USA, Australia, and Kazakhstan. The results (N=132 comparisons) across both agents reveal a significant performance divide contingent on context. For global products the agents achieved 88.9% accuracy (p < 0.0001 against chance). For local products (Sweden), accuracy dropped to 59.5% (p = 0.29), rendering the AI's guidance statistically indistinguishable from random guessing. These findings indicate a cross-market bias consistent with uneven training-data coverage, raising concerns about trust, equity, and accountability, particularly when nutritional judgment shifts from auditable public labels to proprietary inference pipelines. We conclude that AI nutrition lens applications are better framed as assistive, educational tools rather than as replacements for regulated labels, and we highlight the need for auditable datasets and evaluation benchmarks aligned with local food ecosystems.**

## I. Introduction

Recent computer vision and vision–language models enable "AI nutrition lenses Apps"—mobile applications that can convert a photo of a packaged product into near-instant nutritional feedback in a supermarket setting.

### *A. UX of Nutritional Labels*

This trend is driven in part by the usability limits of regulated nutritional labels [1]. For example, in the EU [2], food information can be printed as small as in x-height (⩾1.2 mm, or ⩾0.9 mm for small packages), which can remain difficult to read under real shopping constraints. In addition, nutrition panels are also frequently placed on curved surfaces or near seams and package folds, often on the back of the product. While some consumers rely on labels for safety-critical constraints (e.g., life-threatening allergies), others seek rapid guidance on nutritional aspects. One of them is added sugar, which the World Health Organization recommends limiting to <10% of total energy intake (ideally <5%) [3] to reduce well-known health risks [4]. The combination of health concerns, label legibility limitations, and the cognitive complexity of product comparisons in the supermarket aisle has fostered a growing app ecosystem of AI-powered "lens" applications and label-analysis tools; in addition, general-purpose vision-capable conversational GPT-based chatbots can also be used to interpret packaging images.

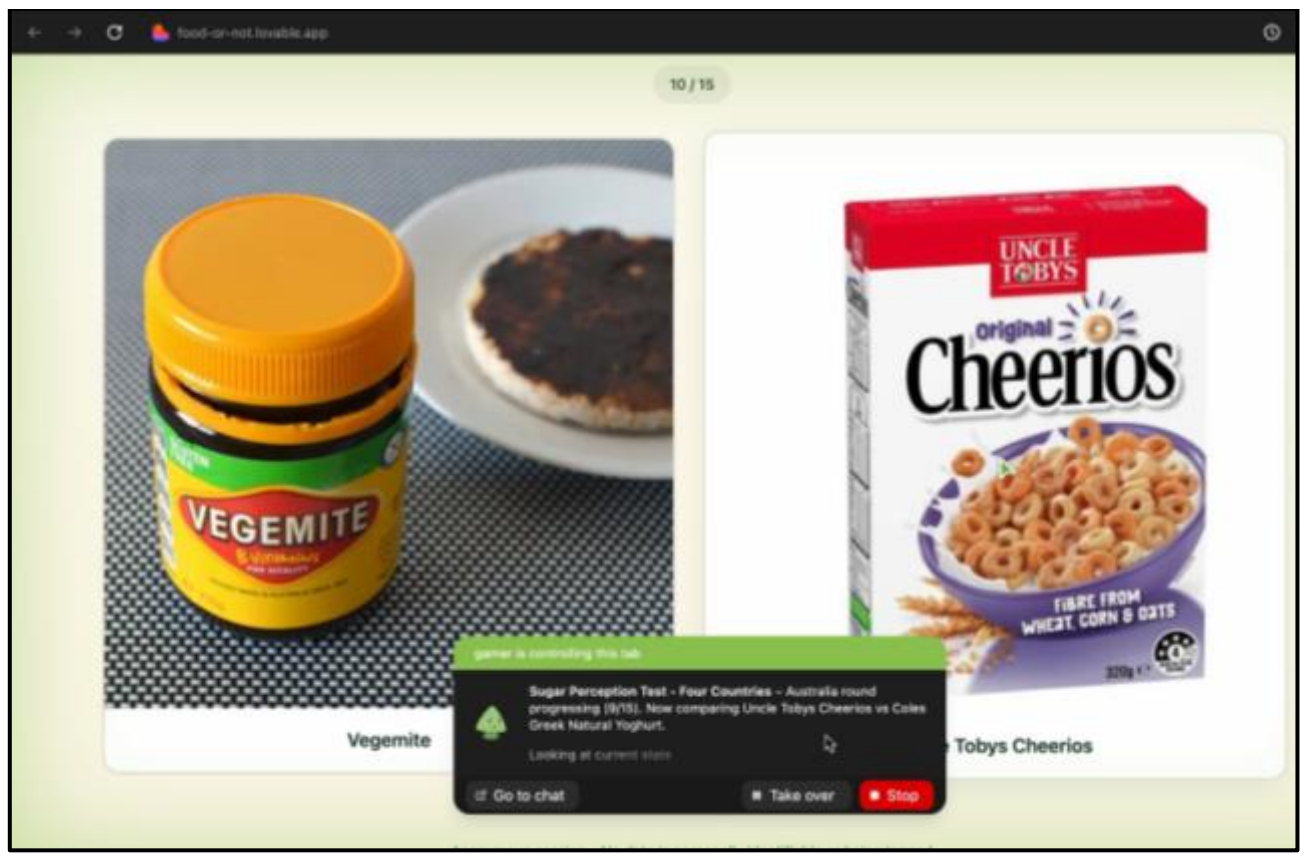

Fig. 1. An agentic browser (see green pop-up) is playing the Two-Alternative Forced Choice game *Food-or-Not [5]*. The goal is to click on the image with less sugar. The aggregates can be used to determine AI-biases across datasets.

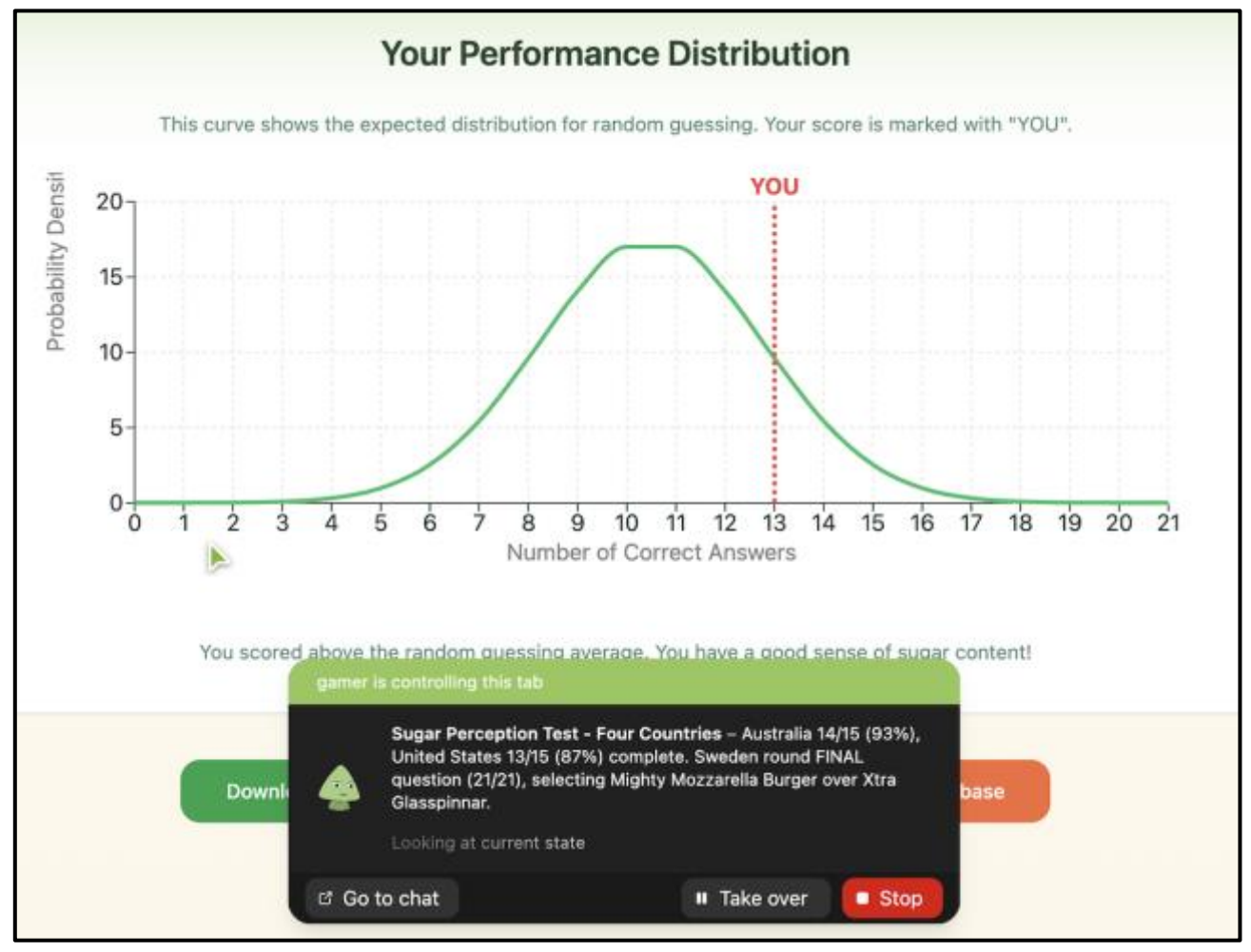

Fig. 2. An agentic browser is in the final screen of the game after completing a set of binary comparisons. In this case, the accuracy of the agent is close to random guessing. Context: Swedish products.

This shift, while convenient for consumers seeking nutritional clarity, raises a governance question: What happens when nutritional information moves from public, auditable disclosures (labels) to private inference pipelines?

### *B. AI-Mediated Nutrition Information*

The trustworthiness of such AI-mediated nutritional inference remains contested. This is not only due to pervasive conflicts of interest with advertisers [6], validation studies show that AI food recognition systems can exhibit systematic inaccuracies in energy and nutrient estimation even when improving convenience or logging workflow [7], [8].

Performance often degrades on culturally diverse foods, indicating dataset-coverage bias and uneven generalization beyond Western-centric product ecosystems [9], [10], [11].

Beyond vision models, large language models can introduce normative "health framing" biases in food evaluation, shaping outputs in ways that may not align with physiological ground truth or local dietary context [12]. Accordingly, reviews of AI in nutrition consistently foreground unresolved concerns around fairness, transparency, accountability, and governance in algorithmic dietary guidance [13], [14], [15].

### C. *Technical State-of-the-Art*

At a technical level, the closest "building blocks" to AI nutrition lenses exist across three partially disconnected strands.

First, strong non-visual baselines demonstrate that added sugar can be inferred from structured nutrition, ingredient, and category data, motivating machine learning as a "digital completion" tool in jurisdictions where added-sugar disclosure is incomplete [16], [17].

Second, multimodal work moves closer to a lens paradigm: vision–language models trained on front-of-pack (FOP) images plus label/ingredient text can estimate multiple nutrient targets and outperform unimodal baselines, demonstrating feasibility of FOP-driven nutrition inference [18].

Third, multiple applied systems primarily function as digital label readers, using object detection, OCR, and parsing to extract nutrition tables and compute scores such as Nutri-Score—effective for digitization, but conceptually distinct from inferring nutrients from packaging cues alone[19], [20]. In parallel, the front-of-pack labeling policy and behavioral literature provide extensive evidence that label formats affect comprehension and choice and offer design principles for labels as regulatory instruments [21], [22].

### D. *Research Gap*

Despite these developments, limited research has directly tested whether AI visual inference from FOP images can substitute for regulated labels under realistic consumer constraints (fast decisions, limited visibility), while also quantifying how performance shifts across global versus local product ecosystems. This study addresses that gap with a constrained binary-game: two AI agents are shown only front-of-pack images and asked to choose which of two products contains less sugar. We chose sugar as a bounded, verifiable proxy for nutrition inference. We evaluate accuracy across four national contexts—Sweden, the United States (USA), Kazakhstan and Australia—to test how product familiarity and data coverage shape AI reliability and to surface implications for trust, equity, and accountability in AI-mediated nutrition guidance.

## II. Methods

### A. *Binary Sugar Game*

We implemented the standard 2-Alternative Forced Choice (2AFC) as a web-based binary-choice task ("Food-or-Not") [5] in which raters viewed pairs of packaged food products using only FOP images and selected which product contained less sugar. The task was delivered through an online public website inspired by similar earlier food games such as '*Scrandle*'. It presents two products per trial and records a forced-choice response (left/right). This format does not 100% approximate a realistic point-of-sale comparison, but it serves effectively the purpose to compare guessing accuracies if the number of data points is sufficient [23]. Fig 1. shows a screenshot of an AI agent playing the game.

### B. *Dataset*

The dataset comprised packaged supermarket items sampled across four national contexts: Sweden (7), the USA (6), Australia (6), and Kazakhstan (6). Only the FOP image for each item was used as input to the task (i.e., no nutrition panel or ingredient list was shown to AI agents). Ground-truth sugar content for each item was obtained from the product's declared nutrition information (e.g., grams of sugar per 100 g/100 ml or an equivalent standardized basis where needed). For each trial, the "correct" answer was defined as the product with lower declared sugar content. Figure 1 shows example products used in the study from the Australia dataset.

**USA**

- Prego Sauce: 8%
- Noosa Yogurt: 16%
- Quaker Real Medleys: 26%
- Clif Bar: 31%
- Kit Kat: 50%
- Honey Smacks Cereal: 59%

**Sweden**

- Köttbullar Vego: 2%
- Festis Orange: 7%
- Yalla Yogurt: 8%
- Caesar Salad: 14%
- Mighty Mozzarella Burger: 17%
- Kanelbulle: 23%
- GlassPinnar: 25%

**Australia**

- Vegemite: 3%
- Weetabix Original: 4%
- Coles Greek Natural Yoghurt: 5%
- Uncle Tobys Cheerios: 15%
- Coles Tomato Ketchup: 29%
- Tim Tam Original: 32%

**Kazakhstan**

- Tan: 0%
- Processed Bread Harrys: 5%
- Activia Yogurt: 7%
- Bun: 20%
- Oreo: 44%
- Nutella: 56%

Images used: https://github.com/orioli/food-or-not/tree/main/src/assets

### C. *AI Agent Systems and Interaction Protocol*

Two AI agent systems were evaluated on the same binary-choice task: Strawberry (agentic browser system) and OpenAI Atlas (vision-capable agent). Each system was instructed to choose which product contained less sugar using only the FOP images, mirroring the human task framing. To reduce contamination from within-session learning, performance was recorded on each system's initial pass through the trial set (as they memorize the correct answers). Figure 2 illustrates OpenAI Atlas interacting with the task interface.

### D. *Outcome Measures and Analysis*

The primary outcome was accuracy, defined as the proportion of trials in which the rater selected the lower-sugar product according to ground truth. Accuracy was computed overall and stratified by national context (Sweden, USA, Kazakhstan, Australia) and by context type (English-language vs. non-English). Because the task is binary, chance performance is 50%; inferential comparisons against chance and between rater types can be conducted using binomial tests and/or contingency-table analyses (e.g., Fisher's exact test), with confidence intervals reported for all accuracy estimates.

Two AI agent systems were shown a website that shows pairs of packaged food products using only front-of-pack images and were asked to identify which contained less sugar. No personally identifiable information was collected, and no sensitive data were involved.

### E. *Ethics*

Ethical approval was granted as exempt by the Institutional Ethics Committee.

## III. Results

The two AI agents (*Strawberry Browser* and *OpenAI Atlas*) were evaluated on their initial pass through on the same task. They showed strongly context-dependent performance rather than a uniform advantage. Table 1 summarizes AI accuracy by national context. For Strawberry agentic browser, accuracy in descending order was: Australia (14/15; 93%), the USA (13/15; 87%), and Kazakhstan (13/15; 87%), and substantially lower in Sweden (13/21; 62%). OpenAI Agentic browser Atlas showed a similar cross-context pattern: Australia (15/15; 100%) and Kazakhstan (15/15; 100%) scored perfect, followed by USA (10/15; 67%) and Sweden (12/21; 57%). Coincidentally, when pooled across the three non-Swedish contexts (Australia, USA, Kazakhstan), both agents achieved the same 40/45 (89%) accuracy; in contrast, performance in Sweden approached chance level for both agents (57, 62%).

To quantify this "local-context drop," we compared each agent's accuracy in Swedish to their individual pooled accuracy on non-Swedish contexts. The reduction was large for both agents (Atlas: 89% to 57%, −32 percentage points; Strawberry: 89% to 62%, −27 percentage points). Consistent with this, binomial tests indicated that performance in Sweden was not distinguishable from chance for Atlas (12/21) and Strawberry (13/21), whereas pooled performance outside Sweden (40/45) was well above chance. A contingency-based comparison of Sweden versus non-Sweden is shown in Figure 3 and Table 2.

Finally, we observed a perfect within-session learning artifact: once an agent had completed the task once, repeated play within the same session could yield near-perfect performance, consistent with memorization of previously seen pairs. This reinforces the importance of evaluating "first exposure" performance when interpreting consumer-facing reliability.

In Table 1, 95% confidence intervals (CIs) for accuracy were computed assuming a binomial distribution of correct/incorrect responses. For each country and AI agent, accuracy was defined as $p= k/n$, where k is the number of correct choices and $n$ the number of trials. Confidence intervals were calculated using the exact Clopper–Pearson method.

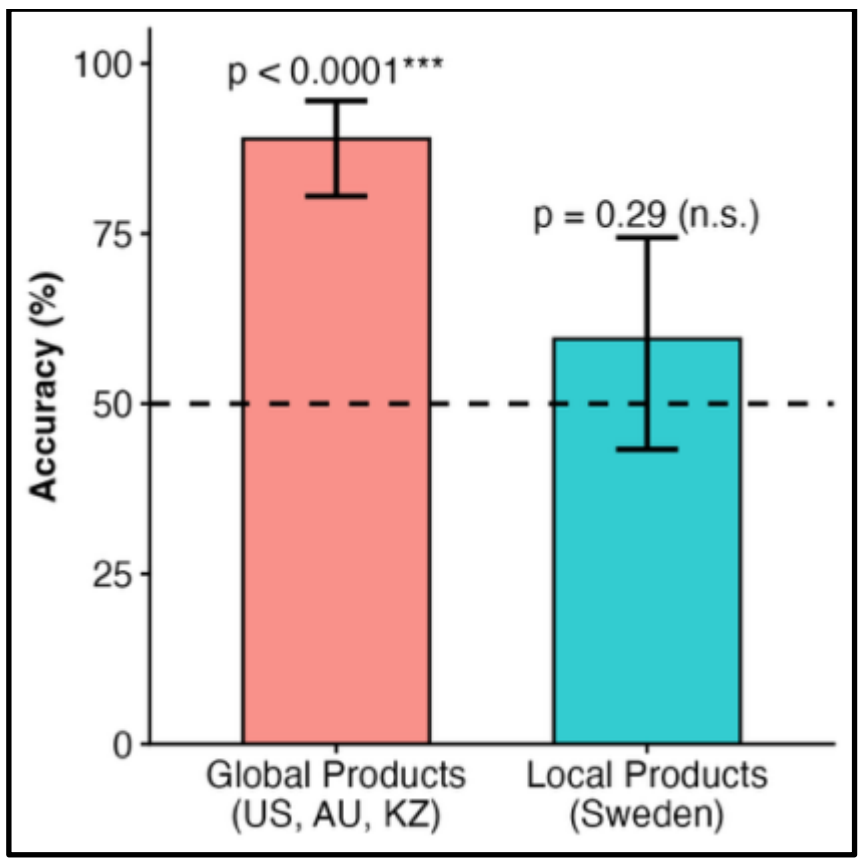

Fig. 3. Pooled performance across global ($n_{trials}$=90) and local ($n_{trials}$=42) contexts for the two AI agents. Error bars indicate 95% CI. The dashed line represents the 50% chance baseline for the 2AFC task. Error bars indicate 95% CI (Clopper–Pearson)

Table I. AI agent accuracy by country (95% CI)

| Country | Strawberry (Swedish) | Atlas (U.S.-made) | Pooled Agents (Combined) |
|---|---|---|---|
| Australia | 93% [80–100] | 100% [78–100] | 97% [83–100] |
| Kazakhstan | 87% [70–100] | 100% [78–100] | 93% [78–99] |
| USA | 87% [70–100] | 67% [43– 91] | 77% [58–90] |
| Sweden | 62% [41– 83] | 57% [36– 78] | **60% [43–74]** |

Table II. Pooled AI agent accuracy by context

| Context | Correct answers / N | Accuracy CI 95% | p-value (vs. chance) |
|---|---|---|---|
| **Global products** (Australia, USA, Kazakhstan) | 80 / 90 | 88.9% [80.5–94.5] | p < 0.0001*** |
| **Local products** (Sweden) | 25 / 42 | 59.5% [43–74.4] | p = 0.29 (Not significant) |

## IV. Discussion

In the global products pairs, AI agents inferred the lower-sugar item with 88.9% accuracy. For reference, this performance is comparable to benchmarks observed for human users in similar tasks. However, a stark disparity emerged in the local, unfamiliar context: when confronted with Swedish products, pooled agent accuracy fell to 59.5%, with the 95% confidence interval overlapping the 50% chance baseline. This drop is fundamentally linked to dataset coverage; the sensitivity to "globality" observed here aligns with established failure modes in vision-recognition systems and reinforces concerns regarding cultural bias in algorithmic food classification.

These results carry significant implications for trust and accountability. Nutritional labels are currently part of a public, auditable infrastructure: their format is regulated, and compliance is legally enforceable. In contrast, AI-mediated judgments are produced by proprietary pipelines whose training sets and decision logics are typically opaque and free from regulatory oversight [24]. When consumers treat AI outputs as authoritative, nutritional judgment shifts from regulated disclosure to private inference, creating an "accountability void". Disparities in performance across markets or languages may remain invisible without systematic auditing, raising urgent questions regarding algorithmic sovereignty.

## V. LIMITATIONS

This study has several limitations that should be considered when interpreting the results. First, the evaluation task was deliberately constrained to a binary (2AFC) judgment of relative sugar content based solely on FOP images. While this design enables clean, verifiable accuracy measurement against ground truth, it does not capture the full complexity of real-world nutritional decision-making, which may involve multi-attribute trade-offs (e.g., fat, salt, portion size) and non-binary choices.

Second, the number of products per national context was modest, and the set was not intended to be statistically representative of each country's full retail ecosystem. As a result, accuracy estimates should be interpreted as indicative; future work should scale the dataset to thousands of products per market, adopting the large-scale paradigms established by benchmarks such as *Food2K*, which encompasses over 1 million images but primarily focuses on ready-to-eat dishes [25].

Third, AI agent performance was evaluated only on first exposure to avoid within-session memorization effects.

Fourth, the mean sugar content of the Swedish products dataset is lower than in the other three countries' data. Finally, the study focused exclusively on sugar content as a bounded nutritional attribute; results may not generalize to other nutrients where visual cues are weaker.

## VI. CONCLUSION

AI 'nutrition lens' systems can provide rapid, image-based guidance, but our cross-context evaluation highlights that accuracy is strongly conditioned on product familiarity and dataset coverage. For globalized, English-language products, general-purpose AI agents achieve high accuracy (88.9%, $p < 0.0001$); however, when products are weakly represented in training data, performance degrades to levels statistically indistinguishable from random guessing (59.5%, $p = 0.29$).

These results suggest that as nutritional judgment shifts from public, auditable labels to proprietary inference pipelines, we risk an 'algorithmic nutrition divide' where guidance quality varies by market and language. Evaluating AI nutrition tools solely on global benchmarks may systematically overestimate their reliability in local retail contexts. We highlight the need for auditable datasets and evaluation benchmarks aligned with local food ecosystems.

## ACKNOWLEDGMENT

Dr. Mei Yen (NUSOM) for insightful discussions, and Lovable AB for free credits.